# Hybridizing WENO implementations of interpolation and reconstruction-wise operation for upwind-biased schemes with free-stream preservation

Qin Li[a, b], Dong Sun[b]

[a] School of Aerospace Engineering, Xiamen University, Xiamen, Fujian, 361102, China,

and

[b] State Key Laboratory of Aerodynamics, Mianyang, Sichuan, 621000, China

**Abstract** Cases have shown that WENO schemes usually behave robustly on problems containing shocks with high pressure ratios when uniformed or smooth grids are present, while nonlinear schemes based on WENO interpolations might relatively be liable to numerical instability. In the meanwhile, the latter have manifested their advantages in computations on grids of bad quality, because the free-stream preservation is easily realized there (X. Deng, et al. J. Comput. Phys. 230 (2011) 1100-1115), and what is more flux-splitting schemes with low dissipations can be engaged inherently as well. Targeting at above dissatisfactions, a method by hybridizing WENO implementations of interpolation and reconstruction-wise operation for upwind-biased schemes with flux splitting employed is proposed and corresponding third-, fifth- and seventh-order upwind-biased schemes are proposed. In addition, the practice establishes an approach to connect two seemingly distinct techniques, i.e. WENO interpolation and reconstruction. Based on the understandings of [Q. Li, et al. Commun. Comput. Phys. 22 (2017) 64-94], the free-stream preservation of proposed schemes is achieved with incorporation of frozen grid metrics in WENO reconstruction-wise operations on split fluxes (T. Nonomura, et al. Computers & Fluids 107 (2015) 242–255). In proposed schemes, flux-splitting schemes with low dissipation can also be applied for the flux on a cell edge. As a byproduct, an implementation of WENO scheme with free-stream preservation is obtained. Numerical examples are provided as following with the third- and fifth-order schemes being tested. In tests of free-stream preservation, the property is achieved as expected (including two implementations of WENO). The computation of 1-D Sod problem shows the capability of proposed schemes on solving ordinary shock discontinuity. 2-D vortex preservation and double Mach reflection are tested on uniformed and randomized grids. The accomplishment by proposed schemes manifests their capability and robustness on solving problems under rigorous circumstances.

## 1 Introduction

Large advances have been achieved on fundamental researches in fluid dynamics since the development and application of high order schemes, where the difference schemes draw much attention due to their mathematical conciseness and implementation efficiency [1, 2]. In spite of the success, it is expected that high order difference schemes can also be applied in engineering complexities, where multi-blocked grids are inevitably used and grids with discontinuous distributions could exist. Among the causes that might affect the robustness and therefore applicability of schemes, it was recognized that the error generated from metric evaluations might play important role in computations [3, 4].

With the awareness of the issue, investigations were carried out and theoretical outcomes were attained, e.g., the methodologies to evaluate the grid metrics and flux derivatives were proposed [5-6] and the central schemes with the nonlinearity introduced through filtering or interpolation were proposed [3, 9-10]. When above two approaches were integrated for usage, it was manifested that errors arisen from metric evaluation could be largely alleviated [7, 8]. In [10], the simulation on flow over a high lift trapezoidal wing was reported with the use of multi-blocked grids, which indicated the capability of the nonlinear high-order central schemes combined with the methodologies in [5-6] for solving practical problems.

As shown in [4], when canonical fifth-order WENO was used without special treatment, large disturbances might be generated on grids with bad quality, and the computation would thereby became either erroneous or unstable. Hence one might think WENO would be less robust when solving problems with grids of bad quality, while nonlinear central schemes with metric cancellation (e.g. WCNS in [8, 10] plus metric evaluations in [5-6]) would be more suitable. However, for problems containing shocks with high ratio of pressure such as double Mach reflection, it was reported [11] that schemes such as WCNS would be more likely to numerical instability on uniformed grids, while WENO schemes felt not difficult to accomplish the computation. These practices seem to lead to the following dilemma: On the one hand, the central schemes with nonlinearity by interpolation such as WCNS could have better performance on deformed grids than flux-based nonlinear schemes such as WENO, on the other hand on smooth or uniformed grids the latter appear more robust than the former.

To improve the robustness of central-type WCNS schemes, the practices have been observed by introducing functions values on cell midpoints together with that on nodes [10-11]. In this study, a different idea is proposed from the perspective of upwind-biased scheme by hybridizing interpolation- and flux-based nonlinear implementations, and through which an approach to connect seemingly distinct techniques, i.e. WENO interpolation and reconstruction-wise operation, is established. As a result, the new nonlinear upwind-biased schemes are derived, and what is more, the property of free-stream preservation (*FSP*) can further be achieved if the metric-evaluation methods in [5-6] are used together with the frozen metric in [19]. Two characteristics of proposed schemes are especially concerned, i.e. the improvement of robustness and the availability of using flux splitting schemes with low dissipation. The details of the above is described in Section 2. In Section 3, typical validating cases are tested to reveal the perspectives of proposed schemes. Conclusions are drawn in Section 4.

## 2 Conservative scheme constructions

Consider 1-D hyperbolic conservation law

$$u_t + f(u)_x = 0. \tag{2.1}$$

If the grid is discretized by $x_i = i \cdot \Delta x$ with $\Delta x$ being the uniform space interval, then at $x_j$

$$(u_t)_j = -\left(\hat{f}_{j+1/2} - \hat{f}_{j-1/2}\right)/\Delta x, \tag{2.2}$$

where $\hat{f}(u)$ is implicitly defined by $f(x) = \frac{1}{\Delta x}\int_{x-\Delta x/2}^{x+\Delta x/2} \hat{f}(x')dx'$. From [1],

$$\hat{f}_{j+1/2} = f_{j+1/2} - \frac{1}{24}\Delta x^2 \left(\frac{\partial^2 f}{\partial x^2}\right)_{j+1/2} + \frac{7}{5760}\Delta x^4 \left(\frac{\partial^4 f}{\partial x^4}\right)_{j+1/2} - \frac{31}{967680}\Delta x^6 \left(\frac{\partial^6 f}{\partial x^6}\right)_{j+1/2} + O(\Delta x^8). \quad (2.3)$$

The task of conservative difference scheme is to approximate $\hat{f}_{j+1/2}$ to a desired order by $h_{j+1/2}$.

Moreover, in order to solve problems with shock waves, shock-capturing capability should be required by means of nonlinear operations. In this regard, at least two typical methodologies are known according to the objects they manipulate with. The first approach employs nonlinear reconstruction based on fluxes on nodes, where the representative ones are WENO and -alike schemes with no fluxes on midpoints. The second approach involves the fluxes at midpoints linearly which is derived by nonlinear WENO interpolation, and the representatives are WCNS and similar schemes. Other than above implementations of nonlinearity, a hybrid technique is proposed in this study, and corresponding third-, fifth- and seventh-order upwind-biased schemes will be introduced next.

2.1 Third-order scheme

The basic methodology to construct upwind-biased schemes is to hybridize fluxes on midpoints and nodes. The practices in [12] have indicated that the excess use of midpoints in upwind manner might incur instability. Hence, on compromising the numerical stability and resolution, only one midpoint $x_{j+1/2}$ is employed here as the basic choice. As the first step, the linear form of the scheme is discussed.

From [12], the third-order upwind-biased scheme has been derived as

$$h_{j+1/2} = \alpha f_{j+1/2} + \left(\tfrac{1}{8}\alpha - \tfrac{1}{6}\right)f_{j-1} + \left(-\tfrac{3}{4}\alpha + \tfrac{5}{6}\right)f_j + \left(-\tfrac{3}{8}\alpha + \tfrac{1}{3}\right)f_{j+1}, \quad (2.4)$$

where $\alpha$ is a free parameter. It is easy to see that when $\alpha = 0$, the linear WENO3 is recovered. Instead of the explicit formulation, Eq. (2.4) can be rearranged in equivalent weighted form as

$$h_{j+1/2} = \sum_{k=0}^{r-1} C_k^r \cdot q_k^r \quad (2.5)$$

where $q_k^r$ is the second-order candidate scheme, $r$ is the number of nodes in $q_k^r$ with $r = 2$ for the third-order scheme, and $C_k^r$ is the normalized linear weight. In order to apply techniques similar to WENO reconstruction shown later, $\{q_k^r\}$ are expected to commonly contain the *same* $\alpha f_{j+1/2}$ as that in Eq. (2.4), and thus

$$\begin{cases} q_k^r = \alpha f_{j+1/2} + \sum_{l=0}^{r-1} a_{k,l}^r f(u_{i-r+k+l+1}) = \alpha f_{j+1/2} + q_k'^r, \text{ and} \\ h_{j+1/2} = \alpha f_{j+1/2} + \sum_{k=0}^{r-1} C_k^r \cdot q_k'^r \end{cases}, \quad (2.6)$$

where $q_k'^r = \sum_{l=0}^{r-1} a_{k,l}^r f(u_{i-r+k+l+1})$. Fortunately, such candidate schemes exist and the coefficients $a_{k,l}^r$ and $C_k^r$ of $\alpha$ at $r=2$ can be derived and shown in Table 1. From the table, it is obvious that the linear WENO3 in weighted form would be recovered by setting $\alpha = 0$, and the weighting formulation seems to be unavailable if $\alpha = 1$.

**Table 1 Coefficients $a_{k,l}^r$ and $C_k^r$ in candidate scheme $q_k^r$**

| r | Coeff. | l or k |   |   |   |
|---|---|---|---|---|---|
|   |   | 0 | 1 | 2 | 3 |
| 2 | $a_{0,l}^r$ | $\frac{1}{2}\alpha - \frac{1}{2}$ | $-\frac{3}{2}\alpha + \frac{3}{2}$ | N/A | N/A |
|   | $a_{1,l}^r$ | $-\frac{1}{2}\alpha + \frac{1}{2}$ | $-\frac{1}{2}\alpha + \frac{1}{2}$ | N/A | N/A |
|   | $C_k^r$ | $\frac{-4+3\alpha}{12(\alpha-1)}$ | $\frac{-8+9\alpha}{12(\alpha-1)}$ | N/A | N/A |
| 3 | $a_{0,l}^r$ | $-\frac{3}{8}\alpha + \frac{1}{3}$ | $\frac{5}{4}\alpha - \frac{7}{6}$ | $-\frac{15}{8}\alpha + \frac{11}{6}$ | N/A |
|   | $a_{1,l}^r$ | $\frac{1}{8}\alpha - \frac{1}{6}$ | $-\frac{3}{4}\alpha + \frac{5}{6}$ | $-\frac{3}{8}\alpha + \frac{1}{3}$ | N/A |
|   | $a_{2,l}^r$ | $-\frac{3}{8}\alpha + \frac{1}{3}$ | $-\frac{3}{4}\alpha + \frac{5}{6}$ | $\frac{1}{8}\alpha - \frac{1}{6}$ | N/A |
|   | $C_k^r$ | $\frac{1}{80}\frac{45\alpha - 64}{9\alpha - 8}$ | $\frac{3}{40}\frac{225\alpha^2 - 494\alpha + 256}{(9\alpha-8)(3\alpha-4)}$ | $\frac{3}{80}\frac{25\alpha - 32}{3\alpha - 4}$ | N/A |
| 4 | $a_{0,l}^r$ | $\frac{5}{16}\alpha - \frac{1}{4}$ | $-\frac{21}{16}\alpha + \frac{13}{12}$ | $\frac{35}{16}\alpha - \frac{23}{12}$ | $-\frac{35}{16}\alpha + \frac{25}{12}$ |
|   | $a_{1,l}^r$ | $-\frac{1}{16}\alpha + \frac{1}{12}$ | $\frac{5}{16}\alpha - \frac{5}{12}$ | $-\frac{15}{16}\alpha + \frac{13}{12}$ | $-\frac{5}{16}\alpha + \frac{1}{4}$ |
|   | $a_{2,l}^r$ | $\frac{1}{16}\alpha - \frac{1}{12}$ | $-\frac{9}{16}\alpha + \frac{7}{12}$ | $-\frac{9}{16}\alpha + \frac{7}{12}$ | $\frac{1}{16}\alpha - \frac{1}{12}$ |
|   | $a_{4,l}^r$ | $-\frac{5}{16}\alpha + \frac{1}{4}$ | $-\frac{15}{16}\alpha + \frac{13}{12}$ | $\frac{5}{16}\alpha - \frac{5}{12}$ | $-\frac{1}{16}\alpha + \frac{1}{12}$ |
|   | $C_k^r$ | $\frac{1}{2240}\frac{175\alpha - 256}{5\alpha - 4}$ | $\frac{1}{2240}\frac{11025\alpha^2 - 24412\alpha + 12288}{(5\alpha-4)(3\alpha-4)}$ | $\frac{3}{2240}\frac{1225\alpha - 1536}{3\alpha - 4}$ | $\frac{1}{2240}\frac{735\alpha - 1024}{3\alpha - 4}$ |

The job left is to determine the appropriate value of the free parameter $\alpha$, and the following considerations are tentatively chosen:

(1) Numerical stability from the view of Fourier transformation. Considering the difference scheme $\delta f_j$ of $(\partial f/\partial x)_j$, its Fourier transformation $\widehat{\delta f_j}$ can be determined analytically.

From [13], some necessary conditions should be satisfied, i.e.

$$d^n \left( \text{Im}(\kappa'(\kappa)) \right) \big/ d\kappa^n \big|_{\kappa=0} \leq 0 \tag{2.7}$$

and

$$\text{Im}(\kappa'(\kappa)) \big|_{\kappa=\pi} \leq 0, \tag{2.8}$$

where $\kappa'$ is the modified scaled wave number with respect to the original scaled one $\kappa$, and $n$ is the minimum order of the derivatives in Eq. (2.7) having non-zero-valued expression. Difference schemes which are worthwhile to check are suggested as the proposed scheme itself and its building candidate(s) ahead of the central one, i.e. Eq. (2.5) and $q_0^2$.

(2) Requirement of convex combinations in Eq. (2.5) or $0 \leq C_k^r \leq 1$.

Considering the above requirements, the confinements of $\alpha$ are found as: $\alpha \leq \frac{4}{3} \cap \alpha \leq 1$ $\cap \{ \alpha \leq \frac{9}{8} \cup \alpha \geq \frac{4}{3} \}$, therefore the intersection set is: $\alpha \leq 8/9$. In this study, a trial is chosen as $\alpha = 13/15$.

Once the difference scheme in linear form is determined, WENO techniques such as reconstruction and interpolation can be introduced, and a nonlinear implementation of Eq. (2.5) is proposed as

$$h_{j+1/2} = \sum_{k=0}^{r-1} \omega_k^r \cdot q_k^r = \alpha f(u_{j+1/2}) + \sum_{k=0}^{r-1} \omega_k^r \cdot q_k'^r \tag{2.9}$$

where $\omega_k$ is the nonlinear weight of $C_k^r$ and smoothness indicators [1, 14], and $u_{j+1/2}$ is derived through WENO interpolation [14]. It is worthwhile to mention that the critical point in above lies in the common component $\alpha f(u_{j+1/2})$ existing in the third-order scheme (Eq. (2.9)) and its second-order building blocks (Eq. (2.6)). Hence by means of Eq. (2.9), the bridge between interpolation-based and flux-based nonlinear implementations is set up, and their respective advantages would be taken use of. Regarding the accuracy relation of Eq. (2.9), a short discussion will be given. First, it is trivial that in Eq. (2.9), the requirement for nonlinear weighting (i.e. $\sum_{k=0}^{r-1} \omega_k^r \cdot q_k^r$) to make $h_{j+1/2}$ achieve the expected order $R$ is still as that of WENO, i.e. $\omega_k^r = C^k \left( 1 + O\left( \Delta x^{R-r_q} \right) \right)$ where $r_q$ is the order of $q_k^r$ ($R=3$ and $r=2$ currently). Next the formulation in right-hand side of Eq. (2.9) makes no change to this requirement considering $\sum_{k=0}^{r-1} \omega_k^r = 1$, therefore the scheme in hybridized form would achieve the designed order providing aforementioned requirement is not violated. Based on above understanding and considering already-made accuracy tests regarding WENO schemes and Eq. (2.4) in [12], similar tests will not be iterated in this study.

In order to derive variables on midpoints, the interpolation should be invoked as

$$u_{j+1/2} = \sum_{k=0}^{r-1} \gamma_k p_k^r \quad \text{and} \quad p_k^r = \sum_{l=0}^{r-1} b_{k,l}^r \cdot u_{j-r+k+l+1}, \tag{2.10}$$

where $b_{k,l}^r$ are coefficients in the candidate interpolation $p_k^r$, and $\gamma_k$ is the nonlinear weight regarding the linear counterpart $D_k^r$. For completeness, the related coefficients are given in Table 2. The nonlinear weights $\omega_k$ and $\gamma_k$ are derived through WENO implementations as

$$\omega_k = \alpha_k^r \Big/ \sum_{l=0}^{r-1} \alpha_k^r \quad \text{and} \quad \gamma_k = \beta_k^r \Big/ \sum_{l=0}^{r-1} \beta_k^r, \tag{2.11}$$

where

$$\alpha_k = C_k^r \Big/ \left(\varepsilon + IS_k^\alpha\right)^2, \beta_k = D_k^r \Big/ \left(\varepsilon + IS_k^\beta\right)^2, \tag{2.12}$$

and where $IS_k^{\alpha,\beta}$ ($IS_k^\alpha$ or $IS_k^\beta$) is the smoothness indicator. The standard ones have forms as

$$IS_k^\alpha = \left(f_{j+k} - f_{j+k-1}\right)^2 \quad \text{and} \quad IS_k^\beta = \left(u_{j+k} - u_{j+k-1}\right)^2. \tag{2.13}$$

In the meanwhile, [12] mentioned that in order to achieve the third order, $IS_k^{\alpha,\beta}$ should adopt counterparts in Eqns. (2.16) and (2.17), i.e., $IS_0^{\alpha,\beta}$ employ the one with the same index while $IS_1^{\alpha,\beta}$ choose $IS_2^{\alpha,\beta}$ in the equations as that in [18]. For convenience, the scheme by Eqns. (2.9)-(2.12) are referred as HWENOIU for "H" denotes hybridizing, "I" denotes WENO implementations of interpolation and reconstruction-wise operation, and "U" denotes upwind-biased characteristics.

As already known, order degradation that originates from smoothness indicators occurs in weighting procedures when critical points are met [15]. If such situation is quite concerned, the technique of piecewise-polynomial mapping function [20] can be employed to map $\omega_k$ and $\gamma_k$, e.g. the one for $\omega_k$ in the third-order scheme is

$$g(\omega) = \begin{cases} C_k \left[1 - \left(\frac{\omega}{C_k} - 1\right)^2\right] & \omega \leq C_k \\ C_k - \frac{1}{C_k - 1}\left(\omega - C_k\right)^2 & \omega > C_k \end{cases}. \tag{2.14}$$

Afterwards, Eq. (2.11) is invoked again to normalize the mapped value and the final weights will be acquired. $\varepsilon$ in Eq. (2.12) is usually $10^{-6}$ when the mapping is absent, and it can be as small as $10^{-40}$ if the mapping is invoked.

At this end, the main part of the third-order HWENOIU scheme except *FSP* property has been described, which actually corresponds to the scheme for the positive flux $f^+$ of $f$. The one for

$f^-$ can be acquired by taking the symmetric form of Eqns. (2.9)-(2.10) with respect to $x_{j+1/2}$. It is conceivable that Eq. (2.9) should be ready for ordinary computations except *FSP* (actually practices by Eq. (2.5) have been shown in [12]; also see computations in Section 3 on uniformed grids). How to acquire *FSP* property will be discussed later in Section 2.3.

**Table 2 Coefficients $b_{k,l}^r$ and $D_k^r$ in interpolation scheme $p_k^r$**

| $r$ | $k$ | $D_k^r$ | $b_{k,l}^r$ | | |
|---|---|---|---|---|---|
| | | | $l=0$ | $l=1$ | $l=2$ |
| 2 | 0 | 1/4 | -1/2 | 3/2 | - |
| | 1 | 3/4 | 1/2 | 1/2 | - |
| 3 | 0 | 1/16 | 3/8 | -10/8 | 15/8 |
| | 1 | 10/16 | -1/8 | 6/8 | 3/8 |
| | 2 | 5/16 | 3/8 | 6/8 | -1/8 |

2.2 Fifth-order scheme

Following the same way, the fifth-order HWENOIU scheme can be derived. Similarly, the linear form of the scheme is discussed first. In [12], the fifth-order upwind-biased scheme with one free parameter $\alpha$ has been derived as

$$h_{j+1/2} = \alpha f_{j+1/2} + \left(\tfrac{1}{30} - \tfrac{3}{128}\alpha\right) f_{j-2} + \left(\tfrac{5}{32}\alpha - \tfrac{13}{60}\right) f_{j-1} + \left(-\tfrac{45}{64}\alpha + \tfrac{47}{60}\right) f_j \\ + \left(-\tfrac{15}{32}\alpha + \tfrac{9}{20}\right) f_{j+1} + \left(\tfrac{5}{128}\alpha - \tfrac{1}{20}\right) f_{j+2}. \quad (2.15)$$

As expected, linear WENO5 will be recovered when $\alpha = 0$. Following the same idea in Section 2.1, the linear weighted form of Eq. (2.15) would be acquired as Eq. (2.5), where the third-order candidate building blocks are still formulated in the form of Eq. (2.6) but with $r = 3$. By means of not complicated deductions, coefficients in schemes such as $a_{k,l}^r$ and $C_k^r$ of $\alpha$ can be derived and are summarized in Table 1 as well. It is apparent that the linear weighted WENO5 will be recovered again by setting $\alpha = 0$.

The next step is to determine the scope of $\alpha$, and the considerations in Section 2.1 are chosen again, i.e. the requirement of numerical stability imposed for Eq. (2.15), $q_0^3$ and $q_1^3$ and the demand of convex combination as $0 \leq C_k^r \leq 1$. The solution of the restrictions are: $\alpha \leq \tfrac{64}{45} \cap \tfrac{4}{5} \leq \alpha \leq \tfrac{20}{21} \cap \alpha \leq \tfrac{4}{3} \cap \{\alpha \leq \tfrac{247}{225} - \tfrac{1}{225}\sqrt{3409} \cup \alpha \geq \tfrac{64}{45}\}$, therefore the intersection set is: $\tfrac{4}{5} \leq \alpha \leq \tfrac{247}{225} - \tfrac{1}{225}\sqrt{3409}$. In this study, a trial of $\alpha = 41/50$ is chosen. It is interesting to note that the first upwind candidate of WENO5, i.e. $q_0^3$ at $\alpha = 0$ is found to be numerically unstable around $\kappa = 0$ because its first non-zero derivative $\left(d^{(4)} q_0^3 / d\kappa^4\right)_{\kappa=0} = 6 > 0$.

Once $\alpha$ is ascertained, the linear part of the weighted scheme is ascertained. Following the procedures in Section 2.1, the nonlinear form of the scheme will be acquired in the form of Eq.

(2.9) with $r=3$. Again, the critical point in the process lies in the common component $\alpha f(u_{j+1/2})$ pertained in the fifth-order scheme and its third-order building blocks. Regarding nonlinear implementation, Eqns. (2.10)-(2.12) are chosen as well to derive fluxes on midpoints and the nonlinear weights $\omega_k$ in Eq. (2.9), and coefficients in $p_k^r$ are shown in Table 2 also. From [1], smoothness indicators of fluxes for reconstruction are

$$\begin{cases} IS_0^\alpha = \tfrac{13}{12}\left(f_{j-2} - 2f_{j-1} + f_j\right)^2 + \tfrac{1}{4}\left(f_{j-2} - 4f_{j-1} + 3f_j\right)^2 \\ IS_1^\alpha = \tfrac{13}{12}\left(f_{j-1} - 2f_j + f_{j+1}\right)^2 + \tfrac{1}{4}\left(f_{j-1} - f_{j+1}\right)^2 \\ IS_2^\alpha = \tfrac{13}{12}\left(f_j - 2f_{j+1} + f_{j+2}\right)^2 + \tfrac{1}{4}\left(3f_j - 4f_{j+1} + 3f_{j+2}\right)^2 \end{cases}, \qquad (2.16)$$

while the ones for interpolation are [14]

$$\begin{cases} IS_0^\beta = \tfrac{1}{3}\left(4u_{j-2}^2 - 19u_{j-2}u_{j-1} + 25u_{j-1}^2 + 11u_{j-2}u_j - 31u_{j-1}u_j + 10u_j^2\right) \\ IS_1^\beta = \tfrac{1}{3}\left(4u_{j-1}^2 - 13u_{j-1}u_j + 13u_j^2 + 5u_{j-1}u_{j+1} - 13u_j u_{j+1} + 4u_{j+1}^2\right) \\ IS_2^\beta = \tfrac{1}{3}\left(10u_j^2 - 31u_j u_{j+1} + 25u_{j+1}^2 + 11u_j u_{j+2} - 19u_{j+1}u_{j+2} + 4u_{j+2}^2\right) \end{cases}, \qquad (2.17)$$

where $u$ can be either primitive or conservative ($Q$) variables. As known, $IS_k^\alpha$ and $IS_k^\beta$ will suffer from order degradation at critical points. When such situation is quite concerned, the similar mapping techniques [20] can be applied by using the function such as

$$g(\omega) = \begin{cases} C_k\left[1 + \left(\tfrac{\omega}{C_k} - 1\right)^3\right] & \omega \leq C_k \\ C_k + \left(\tfrac{1}{C_k - 1}\right)^2 \left(\omega - C_k\right)^3 & \omega > C_k \end{cases}. \qquad (2.18)$$

For completeness, the seventh-order scheme is derived and its coefficients are shown in Table 1 as well. Following the similar procedure as before, the scope of $\alpha$ can be found as: $16/21 < \alpha < \left(12206 - 2\sqrt{3377809}\right)/11025$. One can see that all suggested scopes of $\alpha$ in HWENOIU schemes are below one, therefore the reconstruction-wise operations on fluxes would contribute to $(1-\alpha) \times f_{j+1/2}$ in Eq. (2.3) as well, which is expected to mitigate the possible instability in the evaluation of $\alpha \times f_{j+1/2}$ by interpolation, especially for cases where high pressure ratios exist.

In [12], the spectral properties of Eqns. (2.4) and (2.15) have been carefully studied at $\alpha = 1$ through Fourier transformation and Approximate Dispersion Relation method (ADR). It is conceivable the spectrums of current HWENOIUs would be analogous to that in [12]. Hence only the properties by Fourier transformation are analyzed and shown in Fig. 1. For comparison, the spectrums of fourth- (CS4) and sixth-order (CS6) central schemes are taken for reference, where the form of CS4 is $f_j' = \tfrac{1}{\Delta x}\left[\tfrac{9}{8}\left(f_{j+1/2} - f_{j-1/2}\right) - \tfrac{1}{24}\left(f_{j+3/2} - f_{j-3/2}\right)\right]$ and that of CS6 is $f_j' = \tfrac{1}{\Delta x}\left[\tfrac{75}{64}\left(f_{j+1/2} - f_{j-1/2}\right) - \tfrac{25}{384}\left(f_{j+3/2} - f_{j-3/2}\right) + \tfrac{3}{640}\left(f_{j+5/2} - f_{j-5/2}\right)\right]$. From the figure, HWENOIUs show larger deviations in dispersion relation from the exact solution than that of the central schemes. This is not strange because the use of variables on nodes by HWENOIU has the zero contribution at the scaled wave number $\pi$. In addition, HWENOIU5 show similar dissipation as that of HWENOIU3 at $\pi$, and this is due to the specific choice of $\alpha$ in HWENOIU5 which is derived under the consideration of stability and thereby incurs relatively larger dissipations. For reference, the dissipation of WENO3 at $\pi$ is -4/3, and that of WENO5 is -16/15.

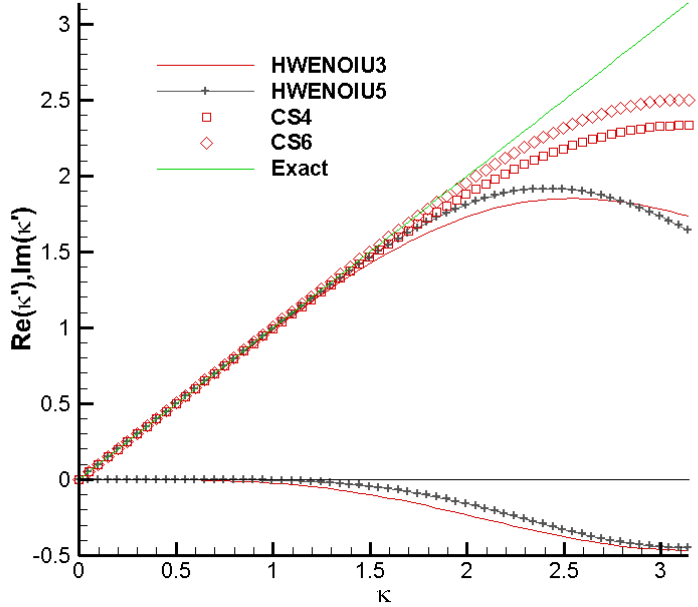

Fig.1 Dispersion and dissipation relations of various difference schemes

2.3 On achieving free-stream preservation

2.3.1 Some reviews and corresponding requirements for HWENOIU

Methodology for linear difference schemes to achieve *FSP* in stationary grids has been extensively investigated [5-8, 16], which regards issues in transformation from physical coordinate system $(x, y, z)$ to computational counterpart $(\xi, \eta, \zeta)$ to acquire conservative equations. In [16], a systematic study was provided for *linear upwind schemes with flux splitting* to achieve the property. In the following, only the implementations are reiterated:

(1) The derivations of grid metrics such as $\hat{\xi}_x$ should employ the conservative or symmetric conservative forms, and based on which analogous conservative or symmetric conservative forms are suggested to compute $J^{-1} = \frac{\partial(x,y,z)}{\partial(\xi,\eta,\zeta)}$. The complete formulations are suggested to [16].

(2) Specific difference scheme should be derived to evaluate grid metrics from the linear upwind scheme as following, i.e. a specific central scheme could be derived for metric evaluation through the process of central scheme decomposition [16]. The central scheme for Eq. (2.4) (i.e. the linear form of Eq. (2.9)) is

$$\frac{\alpha}{\Delta x}\left(f_{j+1/2} - f_{j-1/2}\right) + \frac{1}{\Delta x}\left[\left(\tfrac{1}{16}\alpha - \tfrac{1}{12}\right)\left(f_{j+2} - f_{j-2}\right) + \left(\tfrac{2}{3} - \tfrac{5}{8}\alpha\right)\left(f_{j+1} - f_{j-1}\right)\right]. \quad (2.19)$$

When $\alpha = 13/15$, Eq. (2.19) becomes

$$\frac{13}{15\Delta x}\left(f_{j+1/2}-f_{j-1/2}\right)+\frac{1}{8\Delta x}\left[-\frac{7}{30}\left(f_{j+2}-f_{j-2}\right)+\left(f_{j+1}-f_{j-1}\right)\right]. \qquad (2.20)$$

Similarly, the central scheme for Eq. (2.15) is

$$\frac{\alpha}{\Delta x}\left(f_{j+1/2}-f_{j-1/2}\right)+\frac{1}{\Delta x}\left[\begin{array}{l}\left(\frac{1}{60}-\frac{3}{256}\alpha\right)\left(f_{j+3}-f_{j-3}\right)+\left(\frac{7}{64}\alpha-\frac{3}{20}\right)\left(f_{j+2}-f_{j-2}\right)+\\ \left(\frac{3}{4}-\frac{175}{256}\alpha\right)\left(f_{j+1}-f_{j-1}\right)\end{array}\right]. \quad (2.21)$$

When $\alpha = 41/50$, Eq. (2.21) becomes

$$\frac{1}{\Delta x}\left(f_{j+1/2}-f_{j-1/2}\right)+\frac{1}{\Delta x}\left[\frac{19}{3840}\left(f_{j+3}-f_{j-3}\right)-\frac{13}{320}\left(f_{j+2}-f_{j-2}\right)+\frac{17}{256}\left(f_{j+1}-f_{j-1}\right)\right]. \quad (2.22)$$

The central scheme for aforementioned seventh-order scheme HWENOIU7 in linear form is:

$$\frac{\alpha}{\Delta x}\left(f_{j+1/2}-f_{j-1/2}\right)+\frac{1}{\Delta x}\left[\begin{array}{l}\left(-\frac{1}{280}+\frac{5}{2048}\alpha\right)\left(f_{j+4}-f_{j-4}\right)+\left(\frac{4}{105}-\frac{27}{1024}\alpha\right)\left(f_{j+3}-f_{j-3}\right)+\\ \left(-\frac{1}{5}+\frac{147}{1024}\alpha\right)\left(f_{j+2}-f_{j-2}\right)+\left(\frac{4}{5}-\frac{735}{1024}\alpha\right)\left(f_{j+1}-f_{j-1}\right)\end{array}\right].$$

Regarding the computations by nonlinear HWENIOU schemes discussed later, it is natural to employ the above central schemes for metric evaluations.

Moreover, when midpoints are involved in the scheme as that in current case, the requirement for interpolation, namely directionally consistent interpolation, should be complied with [16].

(3) The flux splitting should follow the following requirement for the linear upwind scheme [16]. Take the flux $\hat{E}$ in $\xi$ direction for example and consider the splitting of fluxes as

$$\hat{E}^{\pm} = \tfrac{1}{2}\left(\hat{E}\pm\hat{A}\cdot Q\right) \quad \text{or} \quad \hat{E}^{\pm} = \tfrac{1}{2}\left(\hat{E}\pm\hat{E}_{ref}\right). \qquad (2.23)$$

In order to achieve *FSP*, $\hat{A}$ and $\hat{E}_{ref}$ in Eq. (2.23) should be locally constant on the dependent grid stencil of upwind operations for both $\hat{E}^+$ and $\hat{E}^-$ when the uniformed-flow condition is imposed [16]. For Eq. (2.4), the range of dependent stencil is ($j$-1, $j$+2), while for Eq. (2.15), the range is ($j$-2, $j$+3). Examples of $\hat{A}$ in Eq. (2.23) like: $\hat{A} = MAX(\max_{k=1...5}|\lambda_k|)\cdot I$ or $\hat{A} = diag(MAX|\lambda_1|,...,MAX|\lambda_5|)$, where $\lambda_i$ is eigenvalue-like variable as $(\lambda_1,...,\lambda_5) = \left(\hat{U},\hat{U},\hat{U},\hat{U}-c|\hat{\xi}|,\hat{U}+c|\hat{\xi}|\right)$ with $c$ denoting sound speed and $|\hat{\xi}| = \sqrt{\hat{\xi}_x^2+\hat{\xi}_y^2+\hat{\xi}_z^2}$, and where *MAX* runs over dependent stencil or the whole $\xi$-direction. For simplicity, the first form of $\hat{A}$ is chosen and *MAX* runs over the whole direction in current study. It is worth mentioning that above requirements are only necessary for the upwind parts in Eqns. (2.4) and (2.15), i.e. the parts except $\alpha f_{j+1/2}$, and there is no restriction for the central part $\alpha f_{j+1/2}$ according to [7-8, 16]. Hence, low dissipation schemes such as AUSMPW+ [17] can be applied for.

2.3.2 On achieving free-stream preservation for HWENIOU with the presence of flux-based, nonlinear operations

Although the above techniques make Eqns. (2.4), (2.15) or (2.5) to achieve *FSP*, they do not entail Eq. (2.9) to achieve the property. The reason is that under the condition of uniformed flow, the nonlinear weights based on fluxes might be different from their linear counterparts due to the presence of non-uniformed grid metrics, and therefore the scheme might deviate from its linear form. Consequently, techniques in previous section will lose the base of their validity and *FSP* would not be achieved. One may wonder that if the linear weights could be recovered from their nonlinear counterparts under the free-stream condition, *FSP* would still be possible for schemes whose nonlinearity is based on fluxes. The straightforward trial from this idea would be the direct use of conservative/primitive variables other than fluxes in smoothness indicators, which resembles that in Eqn. (2.17). Unfortunately, one can find that except in computations of free-stream preservation (Section 3.2) where the uniformed flow is imposed initially, the methodology does not work in practical computations. The consequence indicates the substitution of fluxes with flow variables to evaluate flow smoothness is impractical in nonlinear operations.

Based on above facts, an analysis is given as following and corresponding practice is taken to make HWENOIU achieve *FSP* where nonlinear flux-based operations are employed. First, it is trivial to re-write Eq. (2.9) as

$$h_{j+1/2} = \sum_{k=0}^{r-1} \omega_k^r \cdot q_k^r = \sum_{k=0}^{r-1} C_k^r \cdot q_k^r + \sum_{k=0}^{r-1} (\omega_k^r - C_k^r) \cdot q_k^r$$

It has been mentioned previously that under the WENO framework and when $\omega_k^r = C^k(1+O(\Delta x^{R-r_q}))$ (where $R$ is order of the weighted scheme and $r_q$ is the order of $q_k^r$), the weighted scheme will be *R*th-order accurate. If $q_k^r$ is replaced by its variation $\left(q_k^r\right)^*$ which satisfies $\left(q_k^r\right)^* = c + O(\Delta x^{r_q})$ with $c$ being one common term for $k$, a new $h_{j+1/2}$ can be constructed as

$$h_{j+1/2}^* = \sum_{k=0}^{r-1} C_k^r \cdot q_k^r + \sum_{k=0}^{r-1} (\omega_k^r - C_k^r) \cdot \left(q_k^r\right)^*$$

It can be derived from the equation that

$$h_{j+1/2} = \sum_{k=0}^{r-1} C_k^r \cdot q_k^r + \sum_{k=0}^{r-1} (\omega_k^r - C_k^r) \cdot c + \sum_{k=0}^{r-1} (\omega_k^r - C_k^r) \cdot O\left(\Delta x^{r_q}\right)$$
$$= \sum_{k=0}^{r-1} C_k^r \cdot q_k^r + O\left(\Delta x^R\right) = \hat{f}_{j+1/2} + O\left(\Delta x^R\right)$$

Therefore $h_{j+1/2}$ can make the weighted scheme achieve $R$ order as well providing aforementioned accuracy relations are satisfied.

For HWENOIU, a specific $\left(q_k^r\right)^*$ of $q_k^r$ can be derived by employing the frozen metric technique in [19], i.e., for split fluxes in $q_k''^r$ of $q_k^r$, all related grid metrics such as $\left(\hat{\xi}_x\right)_{j+l'}$ on $j+l'$ point should use one common value $\left(\hat{\xi}_x\right)^*$, e.g. $\left(\hat{\xi}_x\right)^* = \left(\hat{\xi}_x\right)_{j+1/2}$ [19]. Considering that $f_{j+1/2}$ in $q_k^r$ already employs metrics like $\left(\hat{\xi}_x\right)_{j+1/2}$ and the nonlinear algorithm in $\left(q_k^r\right)^*$ keeps unchanged as that in $q_k^r$, it would be seen that under the frozen metric operation, $\left(q_k^r\right)^* = f_{j+1/2} + O\left(\Delta x^2\right)$ for the third-order and $\left(q_k^r\right)^* = f_{j+1/2} - \frac{1}{24}\Delta x^2 \left(\frac{\partial^2 f}{\partial x^2}\right)_{j+1/2} + O\left(\Delta x^3\right)$ for the fifth-order cases. Hence $h_{j+1/2}$ can achieve the formal $R$th-order. Besides, the flux-spitting scheme in $\left(q_k^r\right)^*$ need not comply with some specific form theoretically although L-F scheme is often employed in purpose of simplicity. Recalling Eqns. (2.5) and (2.9), $h_{j+1/2}$ can be further formulated as

$$h_{j+1/2} = \alpha f_{j+1/2} + \sum_{k=0}^{r-1} C_k^r \cdot q_k''^r + \sum_{k=0}^{r-1} (\omega_k^r - C_k^r) \cdot (q_k''^r)^*, \qquad (2.24)$$

where $(q_k''^r)^*$ originates from $q_k''^r$ by similarly invoking aforementioned metric frozen technique [19] for fluxes. At this end, the final form of HWENOIU is accomplished.

When free-stream condition is imposed, the third term in RHS of Eq. (2.24) will be zero because of the null value of $(q_k''^r)^*$. From the previous discussion regarding *FSP* and detailed analyses together with numerical validations in [16] and [12], the principle of HWENOIU to achieve *FSP* is manifested.

From Eq. (2.24), the following flexibility and benefit potential are expected as:

(1) The nonlinear interpolation and reconstruction-wise operation coexist through $\alpha$ in spite of the extra cost of computation to achieve *FSP*, which indicates the flexibility to transfer between two different types of nonlinear operations. Especially when $\alpha = 0$, WENO-version schemes are derived with the achievement of *FSP*.

(2) Although requirements are needed on flux splitting scheme for fluxes on nodes, there is no theoretical restriction on $f_{j+1/2}^{\pm}$. Hence up-to-date schemes with low dissipation can be applied for.

(3) As mentioned in introduction, the flux-based nonlinearity might be more robust in shocks with high-pressure ratios than nonlinear interpolation, therefore the hybrid form in Eq. (2.24) is expected to take the advantages of both implementations. The later validating tests confirm the assumption.

2.4 Implementation summary and discussions

In order to facilitate coding, a summary is made on the implementation of HWENOIU:

(1) Given one HWENIOU scheme with certain order by Eq. (2.24), corresponding linear form (Eq. (2.5)) is ascertained, e.g. Eq. (2.4) or Eq. (2.15) for the third- or fifth-order schemes respectively, and the corresponding central scheme can be defined accordingly (e.g. Eq. (2.19) or Eq. (2.21)). Using the derived central scheme, the grid metrics and Jacobian in conservative and symmetric forms will be evaluated. Regarding coordinates at midpoints, the consistent linear interpolation used in each grid direction with appropriate order will be invoked. For example, the fourth-order interpolation can be chosen for HWENOIU3 as $\phi_{j+1/2} \approx \frac{1}{16}\left(-\phi_{j-1} + 9\phi_j + 9\phi_{j+1} - \phi_{j+2}\right)$; for HWENOIU5, the sixth-order one can be used as $\phi_{j+1/2} \approx \frac{1}{256}\left(3\phi_{j-2} - 25\phi_{j-1} + 150\phi_j + 150\phi_{j+1} - 25\phi_{j+2} + 3\phi_{j+3}\right)$. The same interpolation will be performed to derive metrics at $j+1/2$ in $f_{j+1/2}^{\pm}$.

(2) Then HWENOIU by Eq. (2.24) is implemented to evaluate the first-order derivatives of convective terms in Euler equations, where fluxes on one midpoint and on nodes are involved and are split. Details regarding the flux splitting and nonlinear operations will be further narrated.

(3) There would be three occurrences of flux splitting in Eq. (2.24). For the splitting in $q_k''^r$, the one described in "(3)" in Section 2.3 should be followed and the simple *L-F* scheme is employed in this study. For the splitting at the midpoint $j+1/2$, theoretically no requirement is complied with; however, in only purpose of method validating and avoiding complexity, the same *L-F* schemes is employed there. For the splitting in $(q_k''^r)^*$, the metric frozen technique should be applied, and the similar form of L-F scheme is chosen as well.

(4) For $f_{j+1/2}^+$, flow variables at the midpoint should be nonlinearly interpolated by Eq. (2.10), while the grid metrics on the same position should be derived from ones on nodes by aforementioned consistent interpolation in each grid direction. Similarly, flow variables in $f_{j+1/2}^-$ are nonlinearly interpolated by symmetric scheme of Eq. (2.10) with respect to $j+1/2$. To alleviate numerical oscillations, the commonly-used characteristic variables and corresponding operations are used in interpolation [12].

(5) Using Eqns. (2.11)-(2.13), the nonlinear weights $\omega_k$ can be derived. It is worth mentioning that the split fluxes used in $IS_k^\alpha$ are the same as that in $q_k''^r$, not the metric-frozen ones in $(q_k''^r)^*$.

Comparing Eq. (2.24) and Eq. (2.6) or Eq. (2.5), it is obvious that HWENIOU will increase computations due to the extra reconstruction-wise operation $\sum_{k=0}^{r-1}(\omega_k^r - C_k^r)\cdot(q_k''^r)^*$. Considering that Eq. (2.6) or (2.5) actually works in canonical problems containing shocks [12], one might wonder if aforementioned reconstruction-wise operation would only be invoked when serious situations occur such as shocks with high pressure ratio. Besides, the similarity between $q_k''^r$ and $(q_k''^r)^*$ indicates the possibility to reduce computation by carefully coding. Such attempt will be explored later and only validation of proposed method is concerned currently.

**3 Numerical examples**

The following examples are tested: 1-D Sod shock tube problem, 2-D vortex preservation and double Mach reflection on uniformed and randomized grids. In 2-D cases, *FSP* is also checked on grids of vortex preservation. In all examples, Euler equations are solved with the employment of PPM methods for interpolations in WENOIU and third-order TVD-RK3 method [1] for temporal discretization. It is worth mentioning that in the case of double Mach reflection, the fifth-order interpolation will inevitably yield a negative pressure somewhere in flow field, therefore similar order degradation of interpolation is applied as that in [9].

First, the grid generations are introduced regarding 2-D cases.

3.1 Grid configurations

(1) Grids of vortex preservation

The domain is a rectangular as [-8, 8]×[8, 8], and the grid number is: 81×81. Two grids are employed, namely the uniformed and randomized grids. The generation of the latter are described in [16] and [12], and the randomization factor is set as large as 0.45 within inner area by indices (10, 72)×(10, 72) and is set as zero in rest boundary region for simplicity. It is worth mentioning that the randomization in *x* and *y* direction is alternative at one grid point, e.g., when *x* coordinate of the point is under randomizing, its *y* coordinate will keep constant and vice versa.

(2) Grids of double Mach reflection

The domain is a rectangular as [0, 1]×[0, 4] with the grid number 481×120. Similarly, the uniformed and randomized grids are chosen respectively. A slightly different randomization is employed here where the coordinates are randomized simultaneously

$$x_{i,j} = -\frac{L_x}{2} + \frac{L_x}{I_{\max}-1}\left[(i-1) + 2A_{i,j}\left(Rand(0,1) - 0.5\right)\right],$$
$$y_{i,j} = -\frac{L_y}{2} + \frac{L_y}{J_{\max}-1}\left[(j-1) + 2A_{i,j}\left(Rand(0,1) - 0.5\right)\right]$$

where *L* denotes the length of the domain, Rand(0, 1) is a random function ranging from 0 to 1 and the randomization factor $A_{ij} = 0.1$ in this case (corresponding to a moderate grid oscillations). Still, the randomization is posed for the inner area with 9 points away from the boundaries, while the grids near boundaries keep uniformed for simplicity. In Fig. 2, the grids in the upper-right corner of the domain is shown to qualitatively visualize the randomization.

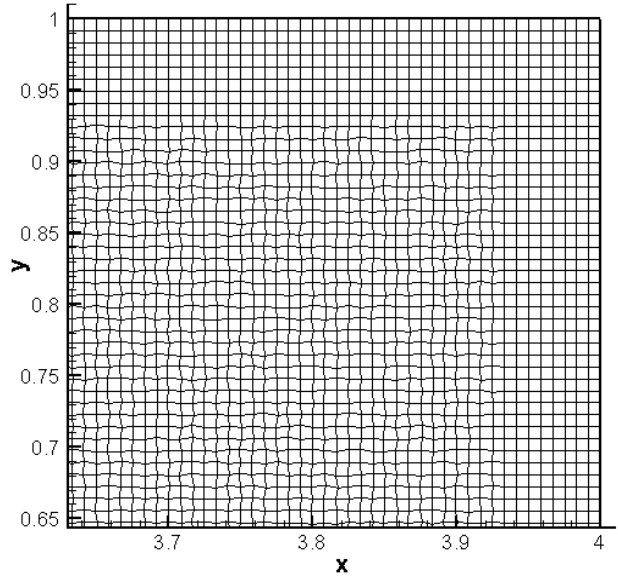

Fig.2 Grids at upper-right corner in double Mach reflection computation.

3.2 1-D Sod problem

The purpose of the test is to check if two rounds of nonlinear operations in Eq. (2.24) would work normally on solving canonical shock wave. The initial condition can be found in [1] with a pressure ratio as 1:10. The computation is advanced to t=2.0 on 100 uniformed grids with $\Delta t$=0.01. The density distributions of HWENOIUs are shown in Figs. 3(a) and 3(b). In the comparison, WENO3-FSP denotes the scheme of Eq. (2.24) at $\alpha$=0 and $r$=2, which is actually equivalent to WENO3 under the uniformed-grid condition, and so does WENO5-FSP similarly. A slight overshoot after the shock and contact discontinuity by WENO3-FSP comes from the use of flux component other than the characteristic variable in reconstructions. From the figure, it can be seen that the hybridization of nonlinear interpolation and reconstruction-wise operation work well on solving 1-D shock problem. As a reminder, in [12] Eq. (2.4) and Eq. (2.15) were also found working well in this problem.

Besides the initial pressure ratio 1:10, we have tentatively tried a ratio up to 1:10000. Under such situation, Eq. (2.4) and Eq. (2.15) fail in the computation while HWENOIU3/5 can still work, which infers that the hybridization of two nonlinearities in Eq. (2.24) would improve robustness of the computation.

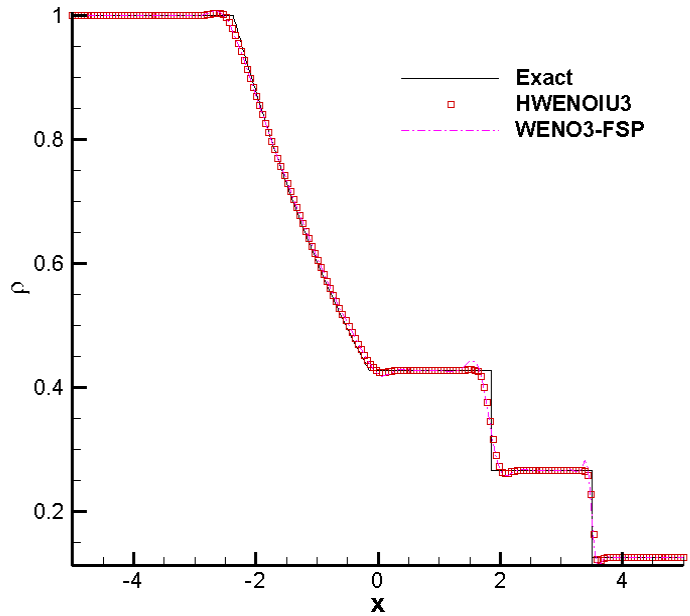

(a) HWENOIU3 and WENO3-FSP.

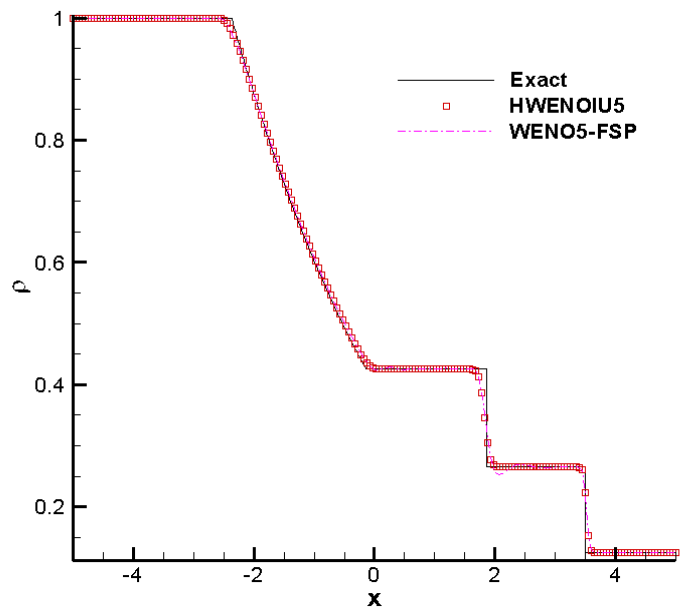

(b) HWENOIU5 and WENO5-FSP.

Fig.3 Density distribution of Sod problem on 100 uniformed grids by tested schemes.

### 3.3 2-D problems

#### 3.3.1 Free-stream preservation

Using randomized grids of vortex preservation described in "(1)" of Section 3.1, the property of *FSP* is checked for HWENIOU schemes. The free-stream condition is:

$\rho_\infty = 1, p_\infty = 1, u_\infty = 1, v_\infty = 0$ and $M_\infty = 1/\sqrt{\gamma}$. The computation advances until $t$=16 with the time step $\Delta t$=0.01. $L_2$ errors of velocity component $u - u_\infty$ and $v$ are shown in Table. 1. The result shows that the achievement of *FSP* is displayed and therefore the algorithm described in Section 2.3.2 works for upwind-biased schemes including nonlinear flux-based operations. As expected, aforementioned WENO-wise implementations, i.e. WENO3/5-FSP, show their achievement of *FSP* as well.

Table 1. $L_2$ errors of $u - u_\infty$ and *v* component in *FSP* test on randomized grids

|  | HWENOIU3 | HWENOIU5 | WENO3-FSP | WENO5-FSP |
|---|---|---|---|---|
| $L_2(u-u_\infty)$ | 9.598E-16 | 1.776E-015 | 5.336E-017 | 1.161E-016 |
| $L_2(v)$ | 2.196E-15 | 2.439E-015 | 3.251E-016 | 8.157E-016 |

3.3.2 Vortex preservation

The problem describes that a vortex initially center-positioned in the domain moves across the right periodic boundary and returns from the left one at *M*=1. The upper and lower boundaries are periodic also. The initial condition of the computation is given in [16]. The computation advances at $\Delta t$=0.01 till *t*=16, which corresponds to one movement circle of the vortex to return to its initial place. The computations are carried out on uniformed and randomized grids respectively, and HWENOIU3, HWENOIU5, WENO3-FSP and WENO5-FSP are tested.

On uniformed grids, the vorticity contours are similar to each other among tested schemes therefore only that of HWENOIU5 are shown Fig. 4 for demonstration. Quantitative comparison is made by drawing the v-component distribution along the horizontally center line of tested schemes in Fig. 5. It can be seen that HWENOIU5 and WENO5-FSP generate almost the same distributions, HWENOIU3 appears a bit dissipative and WENO3-FSP shows the relatively most dissipative.

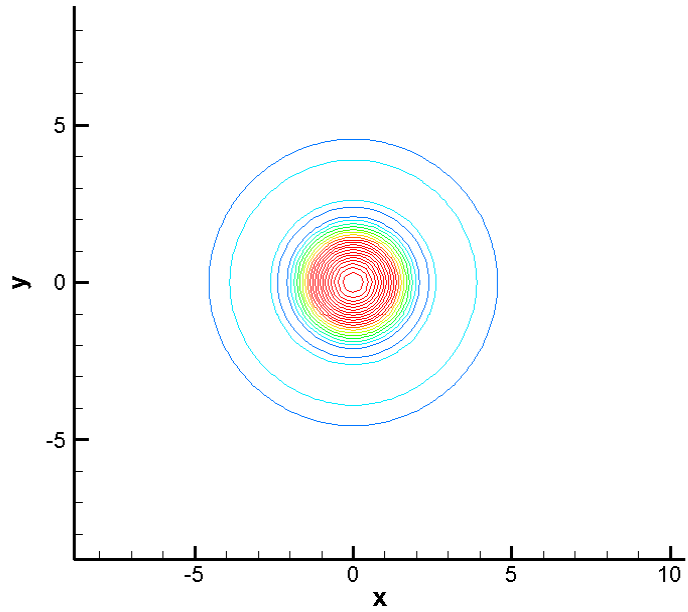

Fig.4 Vorticity contours of vortex preservation problem by HWENOIU5 on uniformed 81×81 grids (Contours from 0 to 0.7 with number 21).

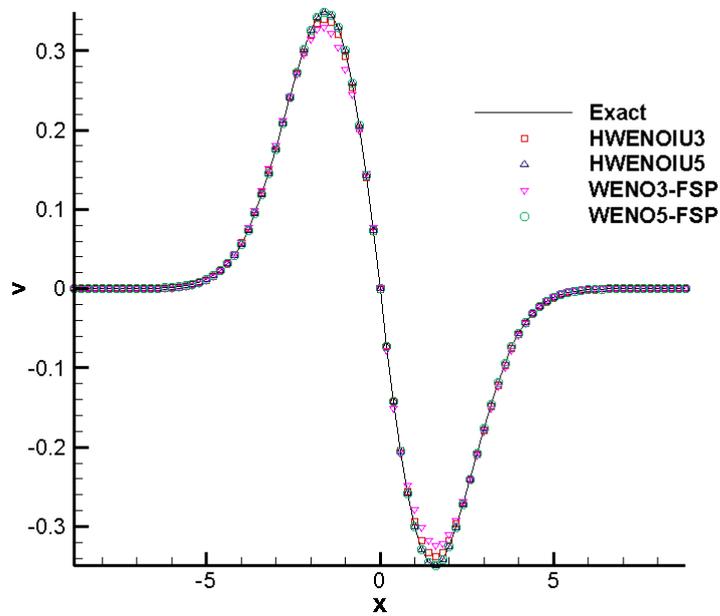

Fig. 5 Distributions of *v*-component along the line at $j = \frac{Jmax}{2} + 1$ of vortex preservation problem on uniformed 81×81 grids by tested schemes

Then the schemes are tested on randomized grids described in "(1)" in Section 3.1, and all schemes have passed the test. On checking, the vorticity contours of HWENOIU3, HWENOIU5 and WENO3-FSP are similar to each other while that of WENO5-FSP appears relatively oscillatory. As a representative, contours of HWENOIU3 are shown in Fig. 6, while that of WENO5-FSP are shown in Fig. 7.

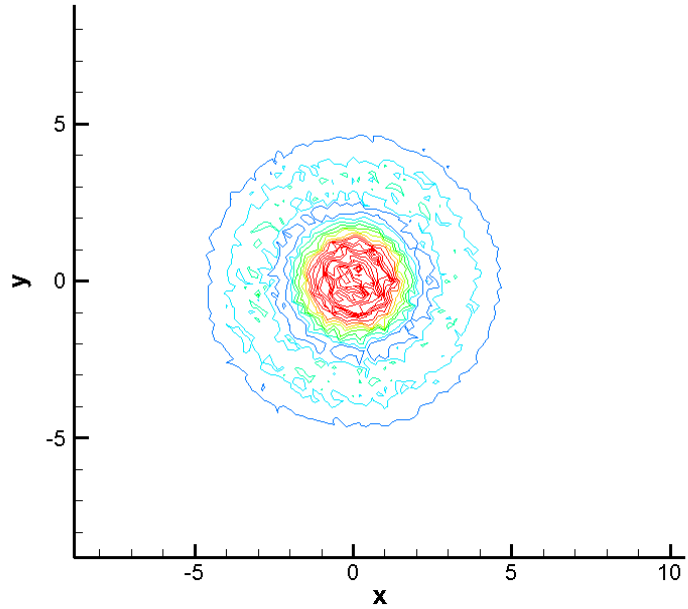

Fig.6 Vorticity contours of vortex preservation problem by HWENOIU3 on uniformed 81×81 grids (Contours from 0 to 0.7 with number 21).

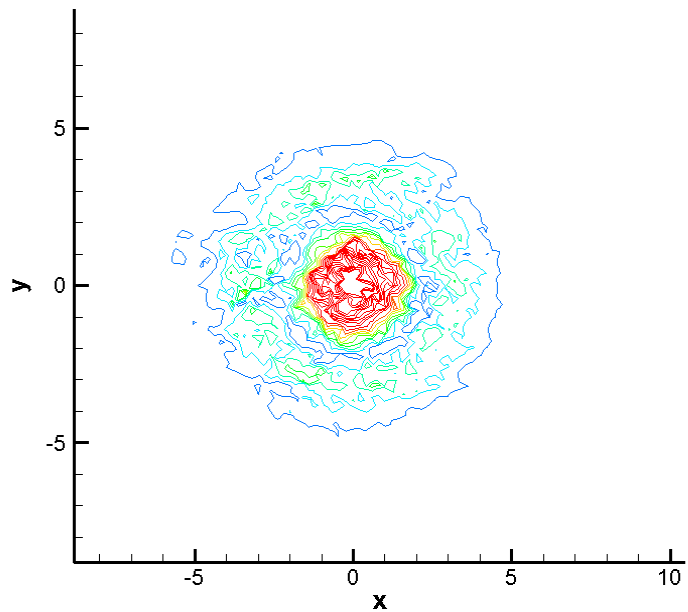

(a) WENO5-FSP

Fig.7 Vorticity contours of vortex preservation problem by WENO5-FSP on uniformed 81×81 grids (Contours from 0 to 0.7 with number 21).

For quantitatively checking, a zoom view of v-component distributions along the horizontally center line of tested schemes is shown in Fig. 8. Comparisons from the figure indicate that HWENOIU5 shows the best agreement with the exact solution, while WENO3-FSP and especially WENO5-FSP appear relatively oscillatory. It is a reminder again that Eq. (2.4) and Eq. (2.15) have also succeeded in this test and the details are suggested to [12].

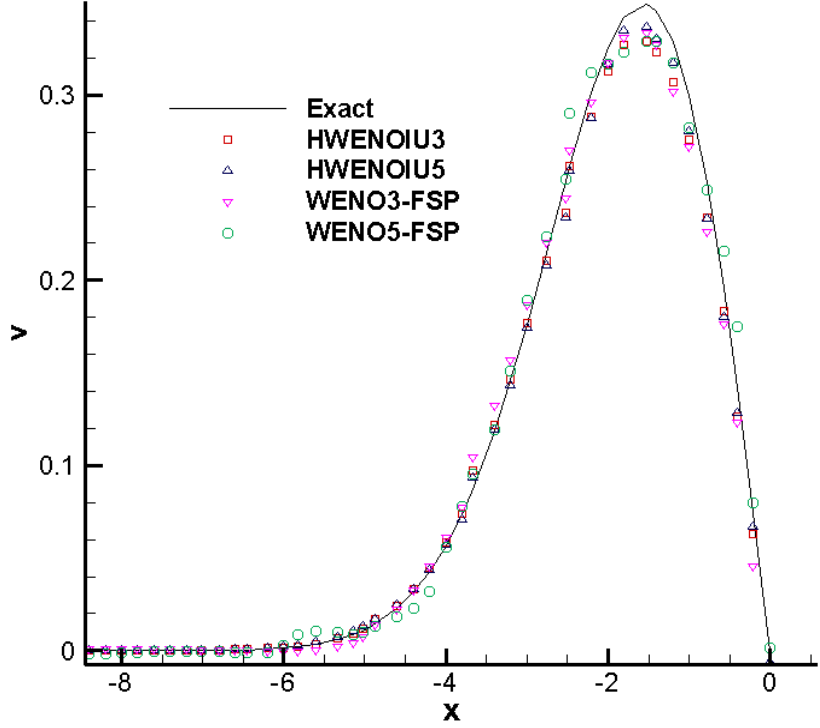

Fig. 8 Zoom view of distributions of *v*-component along the line at $j = \frac{Jmax}{2} + 1$ of vortex preservation problem on randomized 81×81 grids by tested schemes

3.3.3 Double Mach reflection

The computational domain is chosen as [0, 4]×[0, 1], where the reflection wall is placed at the bottom of the domain starting from *x*=1/6. The problem describes a right-moving Mach 10 shock initially with the foot at {*x*=1/6, *y*=0} and with the declining angle 60° to the *x*-axis. The pressure ratio across the shock is 1:116.5 while the density ratio is 1:5.714. Hence the problem denotes a hypersonic flow with high pressure ratio. The computation runs up to *t*=0.2 with Δ*t*= 0.0001, and schemes HWENOIU3, HWENOIU5, Eq. (2.4), Eq. (2.15), WENO3-FSP and WENO5-FSP are tested. As mentioned previously, under uniformed grids WENO3-FSP and WENO5-FSP are respectively equivalent to canonical WENO schemes based on component-wise reconstructions.

All schemes have fulfilled the computation on uniformed grids. The density contours are chosen to show the performances. On checking, the results of HWENOIU3, scheme of Eq. (2.4) and WENO3-FSP are quite similar to each other and that of HWENOIU3 is shown in Fig. 9(a) for representative. On similar consideration, the results of HWENOIU5 is shown in Fig. 9(b) on behalf of WENO5-FSP. The result of Eq. (2.15) appears less oscillatory among fifth-order schemes and is shown individually in Fig. 9(c). It is distinct the fifth-order schemes demonstrate improved resolutions on flow structures.

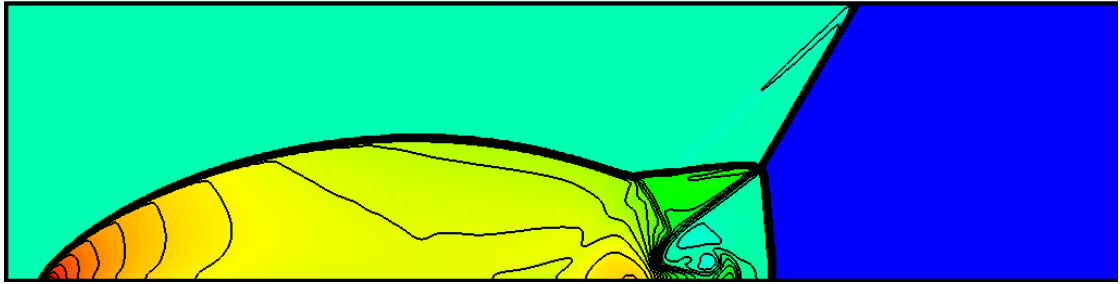

(a) HWENOIU3

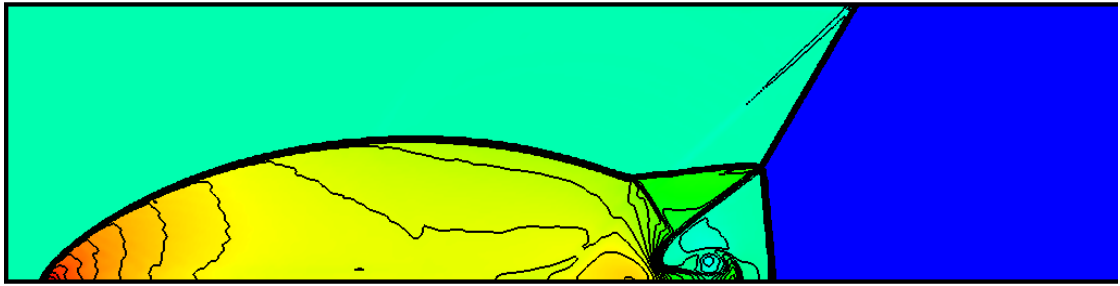

(b) HWENOIU5

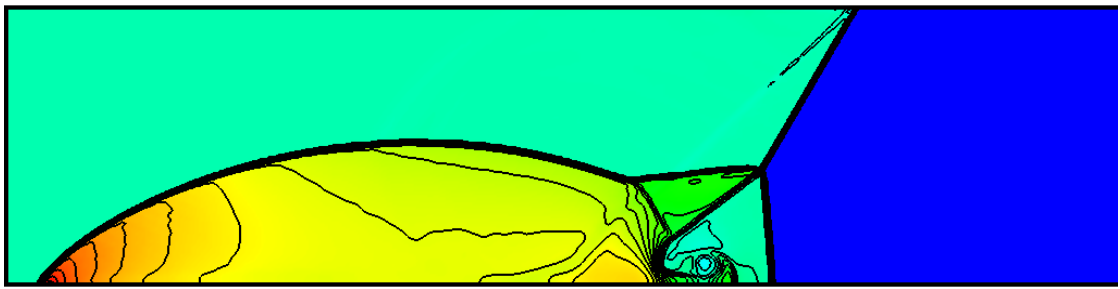

(c) Eq. (2.15)

Fig. 9. Density contours for double Mach reflection on uniformed 481×120 grids at t=0.2 by HWENOIU3, HWENOIU5 and Eq. (2.15) schemes

Then the computations on randomized grids are tested. It turns out that Eq. (2.4) and Eq. (2.15) fail in the computation and blow-ups occur even if the time step is largely decreased. The rest schemes have passed the test and the results are shown in Fig. 10. On checking, HWENOIU3 behaves most robustly and has smoother contours than that of HWENOIU5. Hence the hybridization of two nonlinearities in Eq. (2.24) acquires improved robustness than that of Eq. (2.4) and Eq. (2.15). Although WENO3-FSP and WENO5-FSP have accomplished the computation, their results appear oscillatory, especially in that of WENO5-FSP. Such phenomenon indicates further analysis would be necessary.

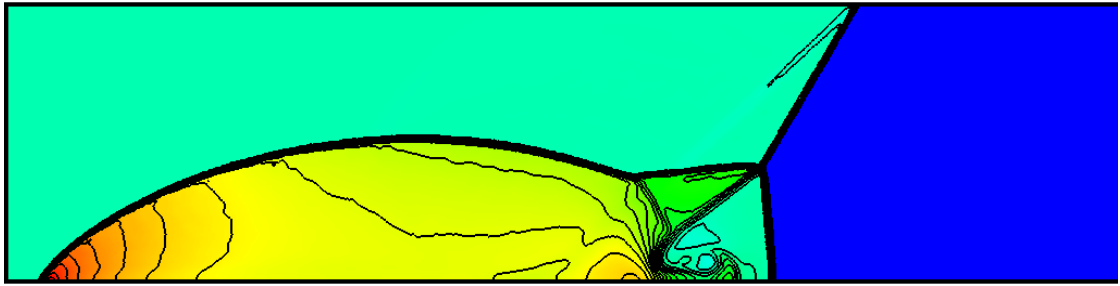
(a) HWENOIU3

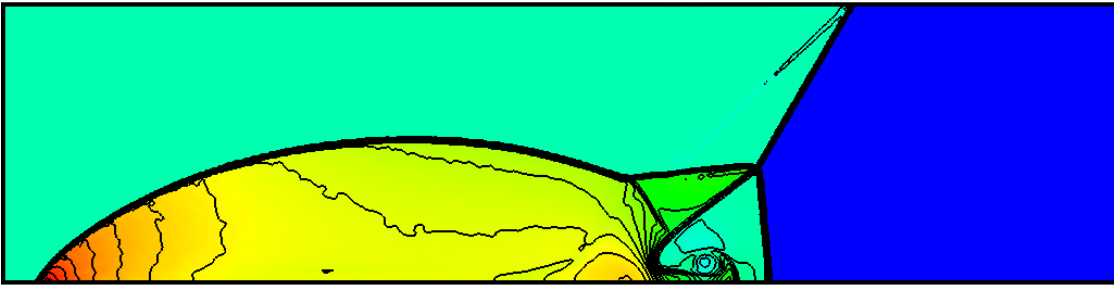
(b) HWENOIU5

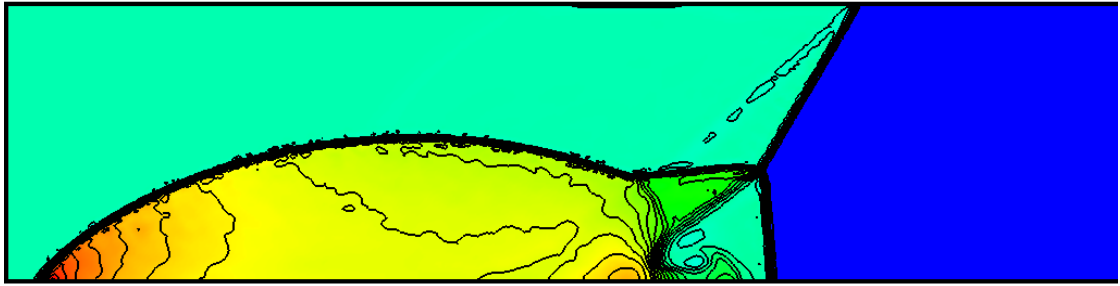
(c) WENO3-FSP

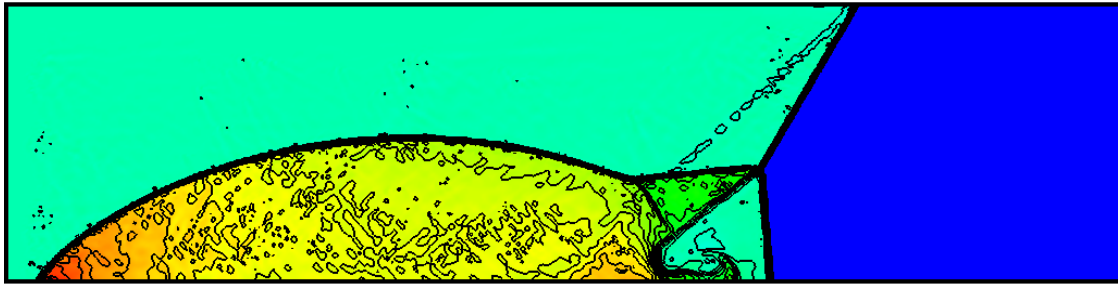
(d) WENO5-FSP

Fig. 10. Density contours for double Mach reflection on randomized 481×120 grids at t=0.2 by HWENOIU3, HWENOIU5, WENO3-FSP and WENO5-FSP schemes

**4 Concluding remarks and discussions**

A hybridization of WENO implementation of interpolation on variables and reconstruction-wise on fluxes are proposed. Through the choice of the parameter $\alpha$, the obtained schemes can transit from the canonical WENO and the hybrid form. Furthermore, the grid metric frozen technique is applied to make the scheme achieve the property of *FSP*. A series of validating cases are tested for proposed schemes. The following concluding remarks are drawn:

(1) Tests of free-stream preservation and vortex preservation on randomized grids validate

the capability of proposed schemes to achieve *FSP*.

(2) The performances of proposed schemes in 1-D and 2-D problems manifest their potential for practical applications, especially on grids with bad quality.

(3) In the case of double Mach reflection on randomized grids, HWENOIU schemes indicate better robustness than their counterparts, i.e. Eq. (2.4) and Eq. (2.15) where nonlinearity only exists in interpolation and the operation on fluxes keeps linear. Such advantage is helpful to simulations in hypersonic flows with shocks of high pressure ratio.

(4) As a byproduct, an implementation of WENO scheme to achieve *FSP* is acquired.

As the price, two rounds of nonlinear operations in HWENOIU schemes will increase computational cost. It is suggested that on the one hand the nonlinear reconstruction-wise operations would only be invoked when necessary, on the other hand careful coding should be done to take full advantage of common computations in candidate schemes (i.e. $q_k''^r$ and $(q_k''^r)^*$ in Eq. (2.24)). However, current study only concerns the validation of the principle and feasibility of HWENIOU, so such exploration would be left for later investigation.

**Acknowledgements.**

The second author thanks for the support of the National Science Foundation of China under the Grant Number 11802324.